\documentstyle[12pt,aasms]{article}
\tightenlines

\def\ni{\noindent}        %No Indent%

\def\hi{\noindent \hangindent=2.5em}
\def\kpc{{\rm\,kpc}}
\def\Mpc{{\rm\,Mpc}}

\def\hnot{{\rm\,km/s/Mpc}}

\def\lsun{{\rm\,L_\odot}}

\def\surfb{{\rm\,mag/arcsec^2}}

\def\aj{{\it A.~J.}, }  %Astronomical Journal%
\def\apj{{\it Ap.~J.}, }  %Astrophysical Journal%
\def\apjs{{\it Ap.~J.~Suppl.}, }  %Astrophysical Journal Supplements%
\def\apjl{{\it Ap.~J.~(Letters)}, } %Astrophysical Journal Letters%
\def\mn{{\it M.N.R.A.S.}, }      %Monthly Notices of the Royal%
\def\aasup{{\it Astr.~Ap.~Suppl.}, }     %A & A Supplements%
\def\spose#1{\hbox to 0pt{#1\hss}}
\def\lta{\mathrel{\spose{\lower 3pt\hbox{$\mathchar"218$}}
     \raise 2.0pt\hbox{$\mathchar"13C$}}}
\def\gta{\mathrel{\spose{\lower 3pt\hbox{$\mathchar"218$}}
     \raise 2.0pt\hbox{$\mathchar"13E$}}}

\def\clock{\count0=\time \divide\count0 by 60
     \count1=\count0 \multiply\count1 by -60 \advance\count1 by \time
     \number\count0:\ifnum\count1<10{0\number\count1}\else\number\count1\fi}

\begin{document}
%\draft

\title{A New Method for Finding\\
	Clusters of Galaxies at $z\gta1$}

\author{Julianne J. Dalcanton\altaffilmark{1,2}}
\affil{Princeton University Observatory,
       Princeton, NJ 08544}
\altaffiltext{1}{email: jd@astro.princeton.edu}
\altaffiltext{2}{present address: Observatories of the Carnegie Institute
of Washington, 813 Santa Barbara Street, Pasadena CA, 91101}

\bigskip
\centerline{\it Submitted to The Astrophysical Journal}
\bigskip

\begin{abstract}

At large redshifts, a cluster or group may be too distant for the
galaxies within the cluster to be detected individually.  However, the
light from these ``undetected'' galaxies still modulates the surface
brightness of the background sky.  Clusters can appear as
$10\arcsec-1.5\arcmin$ sized fluctuations in the surface brightness of
the EBL.  The fluctuations have central surface brightnesses between
roughly 26 and 28 $\surfb$ (in $V$) for clusters between $z=1$ and
$z=2$, and are brighter than the fluctuations produced by background
field galaxies.  While such low surface brightnesses are difficult to
achieve with direct high-resolution imaging, we demonstrate that they
are easily reached in short exposures through smoothing the sky in very
flat CCD images.

For a reasonable extrapolation of the properties and space densities
of clusters and groups, we find that for a wide range of cosmological
assumptions there should be tens of clusters per square degree
visible in the extragalactic background.  The detection rate can range
between 0.1 and 100 clusters per square degree, for extreme
assumptions about the rate of cluster and galaxy evolution.
Unfortunately, the effects of galaxy luminosity evolution and cluster
mass evolution cannot be easily separated; this limits the current
usefulness of this method in discriminating between different
cosmological models.  Drift scans provide sufficient accuracy of
flat-fielding to make the method discussed in this paper an efficient
techinque for finding candidate high redshift clusters.

\end{abstract}

%----------------------------------------------
\section{Introduction}

When we observe the night sky, our attention is immediately and
understandably drawn to what we can easily see, namely stars and
galaxies.  We study these high surface brightness, high contrast
objects in great detail, and obtain deeper and deeper images to make
ever more distant and fainter objects appear above the brightness and
noise of the background sky.  But what about the sky itself?  It is
not a simple uniform background; it contains information about all of
the distant galaxies that are too faint or too low surface brightness
to stand out above the mean rms noise.  The light from undetected
galaxies does not simply disappear.  Instead, the structure of the
background sky must be shaped by the structure of the undetected
galaxies.

Extremely deep CDD images show that faint galaxies almost uniformly
cover the sky (Tyson 1988), so to first order, the approximation that
the sky is simply uniform is a good one.  The large fraction of
galaxies in the field today suggests that the approximately uniform
background of galaxies is primarily made of field galaxies,
distributed along walls and/or filaments.  However, there are also
regions of the universe where the distribution of galaxies is far from
uniform, namely in rich groups and clusters, where the space density
of galaxies is up to factors of $10^4$ times greater than in the
field.  These pockets of galaxies are in the background sky as well,
and while they may be so distant that none of the individual galaxies
may be detected, the integrated light from the cluster is still
luminous and compact, and will leave a clear signature superimposed
upon the extragalactic background light.  The bright patch that the
cluster imprints on the sky may be detected as a specific feature in
the EBL, provided that that cluster is of sufficient richness to be
``rare'' and thus not cover the sky uniformly, and provided that the
surface brightness of cluster rises above random fluctuations
in the EBL on the same angular scale.  By using these bright spots
in the background sky to trace the population of clusters and rich
groups, we can easily identify candidate clusters for further study.

The idea that the extragalactic background can be used to glean
information about the underlying distribution of galaxies is not a new
one.  Schectman (1973,1974) first demonstrated that the power spectrum
of fluctuations in the background light is shaped by the power
spectrum and redshift distribution of the background field galaxies,
and measured the power spectrum from photographic plates.  Martin \&
Bowyer (1989) repeated much of this analysis for the far-UV, where the
sky is much darker.  More recently, Cole, Treyer, \& Silk (1992)
updated Schectman's work, using current determinations of the faint
galaxy correlation function and redshift distribution to predict the
power spectrum of the EBL as a function of wavelength for several
cosmological scenarios.  All of this work has shown that statistics of
the EBL can be a powerful discriminant between different cosmogonies.
However, the EBL can be used as more than a broad statistical tool; it
can be used as a tracer of specific features in the galaxy
distribution.  As can be done with nearby bright galaxies, we can not
only measure the global correlation function, but can identify the
regions where the local correlation is the greatest, namely groups and
clusters.

Using bumps in the EBL to find candidate clusters presents an
extrememly efficient strategy for identifying clusters at high
redshift, especially when compared with the cost in telescope time of
identifying all the cluster galaxies individually.  Surveys that
attempt to find distant clusters through locating overdensities in
galaxy number counts must typically image at high resolution and must
reach at least the characteristic apparent magnitude of the cluster
galaxies ($m^*$ $\approx24$ in $V$ for $z\gta 1$).  This limits
surveys to small areas unless the observer (and Time Allocation
Committee) are willing to invest an extremely large amount of
telescope time. Furthermore, imaging more deeply than $m^*$ does
little to increase the contrast of the cluster against the background
galaxies, given that the slope of the faint galaxy number counts is
typically steeper than the faint-end slope of the cluster luminosity
function.

Instead, searching for large scale fluctuations in
the background sky requires neither high resolution nor particularily
deep data.  By smoothing the sky on the scale of the clusters
(typically tens of arcseconds), one rapidly gains signal-to-noise at
the expense of unnecessary spatial resolution, allowing even short
exposures to reveal very low surface brightness fluctuations.  The
only stringent requirement of this method is extremely accurate
flatfielding.  With the increasing stability of CCD's, and the
coming-of-age of ``shift-and-stare'' supersky flats and transit scans,
this is a diminishing limitation.  For example, an analysis of actual
transit scan data shows that one can survey an area of $\approx10$
square degrees in a single night on a 4m-class telescope, and detect
fluctuations with central surface brightness of $\lta28$
mag/arcsec$^2$ (Dalcanton 1995) by smoothing the background over large
areas.  Furthermore, clusters identified as bumps in the EBL are
almost guaranteed to be at moderate to high redshifts; an
absence of foreground galaxies implies that all of the cluster
galaxies are far enough away to have dropped below the magnitude limit
of the image.

Our goals for this paper are three-fold.  First, we wish to show that
clusters truly appear as easily detected fluctuations in the EBL.  We
do this in \S\ref{coma} by considering how a rich cluster like Coma
would appear if it were at large redshift, and by actually detecting
both simulated and real high-redshift clusters.  Secondly, we would
like to demonstrate that the number of fluctuations due to clusters is
likely to be large enough to make surveying for clusters in the EBL
feasible.  We do so in \S\ref{theory} by calculating
$N(\Sigma,\theta)$, the number density of clusters on the sky as a
function of apparent central surface brightness $\Sigma$ and apparent
angular size $\theta$.  We calculate $N(\Sigma,\theta)$ by
extrapolating the luminosity and surface brightness distributions of
present day clusters backwards to higher redshift, taking into account
the growth of the cluster mass with time and the luminosity evolution
of the galaxies within the cluster for different cosmological
scenarios, and we discuss the resulting distributions in
\S\ref{results} and
\S\ref{results2}.  Finally, we wish to discuss how a survey might
actually be carried out.  We do this in \S\ref{applymethod}, where we
discuss the merits and drawbacks of various strategies, ways to
optimize the search for the most distant clusters, and other possible
sources of background fluctuations.

%----------------------------------------------
\section{The Appearance of Clusters at $z=1-2$}		\label{coma}

First, let us show that it is possible to detect a high-redshift
cluster, even when the galaxies within it cannot be detected
individually.  The necessary ingredient is the that the threshold for
detecting objects of a given surface brightness is a strong function
of angular size.  For example, consider a sky-limited image of an extended
object (i.e.\ a galaxy or cluster) with a mean surface brightness
$\Sigma$ within an angle $\theta$, and a sky surface brightness of
$s$.  The signal-to-noise of this object is
$\Sigma\pi\theta^2/\sqrt{s\pi\theta^2}$ which is proportional to
$\theta$.  Turning this around, at a given limiting signal-to-noise,
the minimum surface brightness that can be detected is proportional to
$1/\theta$; larger objects may be detected at a fainter surface
brightness than smaller objects.  This limiting surface brightness as
a function of angular size is plotted as the dotted line in
Figure~\ref{detectabilitypic}.  Also plotted are the typical mean
surface brightness profiles of a CD galaxy and of a Coma-like cluster
as they would be observed at $z\approx1.5$ (neglecting any luminosity
evolution).  Because of its small angular size, the galaxy never rises
above the threshold for detection.  The cluster, however, can be
detected for large enough apertures, because it has a mean surface
brightness within an angular size $\theta$ that falls off more slowly
with radius than does the limiting surface brightness.

Next, let us look at the properties of the fluctuations that would be
caused by a Coma-like cluster at high redshift.  As an approximation,
we will consider a cluster with a Hubble profile surface brightness
distribution ($\Sigma(r)=\Sigma_0/[1+r/r_c]^2$), a core radius
$r_c=0.1\Mpc$, and an Abell luminosity of $100L^*_V$.  This cluster
would have a central surface brightness of
$\mu_0\approx26,27,28\surfb$, if it were at redshifts of $z=1,1.5,2$,
respectively (ignoring any evolutionary effects or k-corrections).
This corresponds to between two percent and two-tenths of a percent of
the sky brightness in $V$.  Such low pixel-to-pixel noise can only be
reached in a long integration on a large telescope; however, since the
cluster is an extended object, we can sacrifice resolution for greater
sensitivity by smoothing over large scales.  The angular size of the
cluster, defined by the radius at which the surface brightness falls
below the limiting isophote, would be roughly $10\arcsec-1.5\arcmin$
for $z=1-2$ and for a limiting isophote $\mu_{lim}=29.0$.  Thus, if
the cluster were observed with a CCD with a pixel scale of
$0.5\arcsec/$pixel, then smoothing the image on the scale of the
cluster would increase the signal-to-noise of the large low-surface
brightness fluctuation by over a factor of 20 from the single pixel
signal-to-noise, provided that the flat-fielding variations are
smaller than the amplitude of the fluctuation produced by the cluster.
Therefore, although the cluster produces a very low surface brightness
fluctuation, by smoothing of the EBL such fluctuations should be
easily detectable.  The angular size of the cluster, which corresponds
to a radius of a few core radii in the frame of the cluster, is much
smaller than typical a typical field of view, although the Abell
radius ($1.5\Mpc$) of a cluster is typically larger.  Because the
detectable portion of the cluster easily fits within most CCD cameras,
the fluctuation is less prone to being confused with large scale
flat-fielding errors.  This is even less of a problem for CCD drift
scans.

Figure~\ref{fakeclusterpic} is a more concrete example, showing how a
rich, high-redshift cluster could be detected in the EBL by merely smoothing
the background light.  In order, the panels show an artificial rich
cluster at $z=1$, the same cluster embedded in a uniform background of
field galaxies, the cluster and the field galaxies plus the EBL, and
finally the result of smoothing the previous image.  Notice how after
smoothing the cluster stands out prominently above the fluctuations in
the background, although it is completely invisible beneath the
poisson noise of the EBL in the unsmoothed image.

Figure~\ref{emssclusterpic} shows a similar sequence for an actual
X-ray selected cluster at z=0.83, generously provided by Gioia
\& Luppino (1994).  The sequence of images show how the cluster
could be detected as a bright patch in the EBL in an exposure of
less than 30 seconds on a two-meter telescope, immediately demonstrating
the feasibility of using the EBL to locate rich high-redshift clusters.
The cluster is $6\sigma$ above the mean sky level, and obviously stands
well above the level of the background fluctuations.  The rms fluctuations are
only 12\% larger than are seen after smoothing comparable
uniform Poisson noise,
suggesting that variations in the distribution of field galaxies are
not resposible for the majority of the fluctuation amplitude at these scales,
especially considering the residual flat-fielding errors.  This is not
surprising in light of the very weak angular correlation of faint galaxies
(Koo \& Szalay 1984, Stevenson et al.\ 1985, Efstathiou et
al.\ 1991, Pritchett \& Infante 1992, Bernstein et al.\ 1993).

%----------------------------------------------
\section{Calculating the Observed Properties of Distant Clusters}\label{theory}

We now turn our attention from detecting an individual cluster in the
EBL towards understanding the properties of the entire population of
clusters that could be revealed in the background light.  Can clusters
be detected with sufficient frequency to make surveying the EBL
practical?  With what range of surface brightnesses do clusters
appear?  What are the richnesses of the systems that produce
fluctuations of a given amplitude and angular scale?  At what
redshifts are the detected clusters?  We will answer these questions
by calculating the apparent size, surface brightness, and number
density on the sky of clusters as a function of their redshift and
luminosity, using a reasonable extrapolation from cluster properties
at low redshift.

First, we wish to calculate the distribution of cluster surface brightnesses.
To constrain the apparent surface brightness of high redshift
clusters, we may capitalize on a fortunate coincidence uncovered
by West, Oemler, \& Dekel (1987,1989).  After reanalyzing data on 29
Abell clusters in a consistant manner, they concluded that the surface
brightness profiles of the clusters in the outer regions ($r\gta0.1\Mpc$)
were well fit by a deVaucouleur's
profile ($
\Sigma_r^\prime(r) =
	\Sigma_e^\prime \exp{-7.67\left[\left(r/R_e\right)^{1/4} - 1\right]}
$)
where $R_e$ is the half-light radius of the cluster, and $\Sigma_e^\prime$
is the intrinsic surface brightness (in units of luminosity per area)
at $R_e$.  Furthermore, West, Oemler, \& Dekel found that $R_e$ and
$L_{tot}$ obey a simple scaling equation:

\begin{equation}
R_e \propto L_{tot}^{0.51\pm0.07}
\end{equation}

\ni ($H_0=100$ km/s/Mpc, $q_0=0.5$), immediately implying that
clusters over a wide range in luminosity ($30\lta L_{tot}/L^* \lta
550$) have approximately {\it constant} characteristic surface
brightness.  (For a deVaucouleur's profile the total integrated
luminosity of the cluster, $L_{tot}$ is $7.22\pi R_e^2
\Sigma_e^\prime$.)  Assuming that galaxies trace the mass of a
cluster, the clusters must have a constant characteristic surface
density as well.  If the scaling relationship between $R_e$ and
$L_{tot}$ holds true with increasing redshift, as we would expect if
the growth of groups and clusters is scale-invariant (Kaiser 1987),
then the characteristic surface density is not only independent of
total mass, but is independent of time as well.  All evolution in the
intrinsic surface brightness is then due only to luminosity evolution
of the galaxies themselves, a process that can be readily modelled
with stellar synthesis codes.  We are thus in the rather surprising position
of being able to estimate the apparent surface brightness of groups
and clusters of galaxies as a function of redshift.  Letting
$\Sigma^\prime$ be the intrinsic central surface brightness and
fitting to the data in West, Oemler, \& Dekel (1989), we take the
probability of a cluster at redshift $z$ having an intrinsic central
surface brightness $\Sigma^\prime$ at to be a log-normal distribution

\begin{equation}				\label{psigprime}
p_{\Sigma^\prime}(\Sigma^\prime,z)\,d\Sigma^\prime =
{1\over{\sqrt{2\pi\sigma^2_\Sigma}}}\,e^{
-{{\left(\ln{\Sigma^\prime} - \ln{\Sigma_0^\prime(z)}\right)^2}\over
{2\,\sigma^2_\Sigma}}}\,{{d\Sigma^\prime}\over{\Sigma^\prime}}
\end{equation}

\ni where

\begin{equation}
{\Sigma_0^\prime(z) = \Sigma_0^\prime\,\Upsilon(z)}
\end{equation}

\ni and where

\begin{equation}				\label{upsilon}
\Upsilon(z) \equiv \left({{M/L_X(0)}\over{M/L_X(z)}}\right)
\end{equation}

\ni absorbs the effects of galaxy spectral evolution and of the change
in the rest wavelengths observed through filter $X$ with increasing
redshift.  We have determined $\Upsilon(z)$ using the stellar
population synthesis code developed by Bruzual \& Charlot (1993).  We
use a galaxy formed at $z=2$ with an exponential star-formation rate
of $\tau=t_{lookback}/2$ as a ``spiral'' galaxy, and use a galaxy
formed at $z=4$ with $\tau=t_{lookback}/14$ as an ``elliptical''
galaxy (see Figure 7 of Bruzual \& Charlot (1993)).

To find the specific values of $\Sigma_0^\prime$ and $\sigma_\Sigma$,
we choose to make the assumption that the centers of the clusters are
better fit by Hubble profiles with core radii than by steeper
deVaucouleur profiles
\footnote{As will be shown later in this paper, only the core of the cluster
is typically detectable in the EBL ($r\lta0.1\Mpc$).  Thus the results
of this paper will be sensitive to the particular form of the profile
chosen to model the core.  Choosing a Hubble profile to model the
center regions is most likely an improvement over choosing to
extrapolate the deVaucouleur profile all the way into the center.}.
We take a core radius $r_c=R_e/9.83$, which preserves the luminosity
within $R_e$, gives a surface brightness profile that differs by no
more than 6\% from the deVaucouleur's between $r_c$ and $R_e$, and
reduces the central surface brightness by a factor of 17.96 (or
arguably more intuitively, by $3.14\surfb$).  Rescaling the
distribution of central surface brightnesses in West, Oemler \& Dekel
by 17.96, we find $\Sigma_0^\prime=1.63\times10^3\, L^*/{\rm Mpc}^2$
and $\sigma_\Sigma=0.55$.  Throughout this paper we use West, Oemler,
\& Dekel's value of $L_V^*= 1.3\times10^{10} h_{100}^{-2}\,\lsun$ and
$M^*_V=-21.0$.

We make a luminosity dependent correction to $\Sigma_0$ to improve the
estimate of the central surface brightness for poorer clusters.
Bahcall's (1980) analysis of the optical properties of Morgan's poor
clusters (Morgan, Kayser \& White 1975, Albert, White, \& Morgan 1977)
shows that the central surface brightness of poor clusters drops by
roughly a factor of 3 from richness class 0 through richness class $-3$,
(which corresponds to 10 galaxies within 1.5 $h^{-1}_{100}$ Mpc of the center
-- see Bahcall's Figure 1).  We include this effect as a correction to
the central surface brightness of clusters with $L_{1.5}<30 \,L^*$.
We neglect clusters (or more truly, groups) with $L_{1.5}<10 \,L^*$,
as their profiles and central surface brightnesses are not well
determined.

While equation~\ref{psigprime} gives the distribution of intrinsic
surface brightnesses, at large redshifts the apparent surface
brightness will be much smaller due to the different redshift
dependences of the luminosity and angular diameter distances.  As we
are interested in the observable properties of clusters, we must
correct equation~\ref{psigprime} to account for this ``cosmological
dimming''.  We may relate the apparent surface brightness of a cluster
at redshift $z$, $\Sigma(z)$, to the intrinsic surface brightness
$\Sigma^\prime(z)$ as

\begin{equation}					\label{sigp2sig}
\Sigma(z) = {{\Sigma^\prime(z)}\over{4\pi\left(1+z\right)^4}}.
\end{equation}

\ni Note the unfortunate fact that both $\Sigma$ and $\Sigma^\prime$ have
the same dimensions, where the former is measured in flux per
sterradian, and the latter is in luminosity per area.  To attempt to
reduce confusion between these two, the primed superscript will be used
for intrinsic surface brightnesses.  With this transformation, the
distribution of apparent surface brightnesses is

\begin{equation}					\label{psigmaz}
p_\Sigma(\Sigma,z)\,d\Sigma =
p_{\Sigma^\prime}(\Sigma^\prime=4\pi(1+z)^4\Sigma,z)\,\,4\pi(1+z)^4\,\,d\Sigma.
\end{equation}

If one instead wishes to express the apparent suface brightness in
terms of $\mu$, in magnitudes/square arcsecond, one may use

\begin{equation}					\label{mufromsigequ}
\mu(z) = M_* + 48.82 - 2.5\log{\left({\Sigma(z)\over{L_*/\Mpc^2/{\rm
radian}^2}}\right)}.
\end{equation}

\ni to change variables in equation~\ref{psigmaz} from $\Sigma$ to
$\mu$.

We now have the distribution of apparent central surface
brightnesses as a function of redshift.  To get the surface density of
clusters on the sky, we need to combine the distribution in
equation~\ref{psigmaz} with the number density of clusters as a
function of redshift and luminosity, $n_L(L,z)dL$.  The surface density
on the sky of clusters at $z$ with luminosity $L$ and apparent central
surface brightness $\Sigma$ is then

\begin{equation}					\label{Nsigmalz}
N_L(\Sigma,L,z)\, d\Sigma \, dL \, dz
	= n_L(L,z)\,dL\,dV(z) \,\times\, p(\Sigma,z)\,d\Sigma
\end{equation}

\ni where

\begin{equation}
dV(z)= \left({{c}\over{H_0}}\right)^3{{\chi^2(z)}\over{(1+z)^3\sqrt{1+2q_0z}}}
\,dz
\end{equation}

\ni is the comoving volume of one sterradian of a shell between $z$ and
$z+\delta z$ and

\begin{equation}
\chi(z) = {1\over{q_0^2}}\left[q_0z + (q_0 - 1)(-1+\sqrt{1+2q_0z})\right].
\end{equation}

We must determine the form of $n_L(L,z)$ to use in
equation~\ref{Nsigmalz}.  We do this by taking the number density of
clusters at $z=0$, $n_L(L)$, then using linear theory to estimate how
the number density might scale with redshift.  Moore, Frenk, \& White (1993)
have analyzed the CfA redshift survey (Davis et al. 1982, Huchra et
al. 1983) to find $n_{L}(L)\,dL$, the comoving density of groups and
clusters with luminosity between $L$ and $L+dL$.  They first use a
friends-of-friends linking algorithm, fine-tuned through comparisons
with simulations, to identify groups and clusters.  They
then scale the luminosity of the groups to include the missing luminosity of
galaxies lying below the magnitude limit of the survey.  They find
that $n(L)$ is well fit by a double power-law:

\begin{equation}			\label{n(L,z=0)}
n_L(L) = n_0 \, \left[ \left({L \over L_0}\right)^{\beta_1} +
                       \left({L \over L_0}\right)^{\beta_2} \right]^{-1},
\end{equation}

\ni with $\beta_1=1.34$, $\beta_2=2.89$, $L_0 = 7L_*$, and
$n_0 = 1.26\times10^{-3} \, h_{100}^3 \Mpc^{-3}$.

While percolation algorithms are a convenient unbiased way to generate
group catalogs, they do not offer a particularily well-defined
definition of the group luminosity.  This makes it difficult to make a
sensible comparison between the group luminosity function and any
other study.  Moore et al.\ argue that their groups are approximately
bounded by equidensity surfaces at 10-100 times the mean galaxy
density.  The best that we can do to attempt to define our group
luminosities in the same manner is to calculate the luminosity within
some equi-surface density countour $\Sigma_{lum}^\prime$.  To determine
$\Sigma_{lum}^\prime$, we use the total luminosity and effective radius
of the deVaucolouer's profile fit to Coma's surface brightness profile
given in West, Oemler, \& Dekel (1987,1989), and calculate the limiting
surface brightness within which one measures the luminosity reported
by Moore et al.  We find $\Sigma_{lum}^\prime=1.4\,h_{100}^2\,L_*/\Mpc^2$.
This value is uncertain by a factor of {\it ten}, given Moore et al.\'s
quoted range in the variation of the bounding equidensity surfaces.
Because of the slow fall-off with luminosity in our assumed
Hubble-law cluster profile, this yields considerable uncertainty
in identifying the correct number density to use for a cluster with
a given luminosity within $\Sigma_{lum}^\prime$.  If we shift our choice of
$\Sigma_{lum}^\prime$ up by a factor of 10, the peak detection rate for
clusters
goes up by $\sim$40\%.  If we shift our choice down by a factor of 10, the
detection rate drops by $\sim$25\%.

We now estimate the $z$-dependence of $n_L$.  In bottom-up
scenarios of structure formation, the masses of galaxies and clusters
continually grow through infall of density pertubations collapsing on
progressively larger scales, or through merging of small clumps of
mass to form larger ones.  For a spectrum of pertubations that has
no preferred scale, the masses of clusters grows self-similarily
with time.  Thus, $n_M(M)$, the number density of objects as a
function of mass $M$, should have the same shape at every
redshift, but should shift towards larger masses with increasing time.

To calculate the rate at which the characteristic cluster mass grows
with time, we use a result from Peebles (1980).  In linear theory,
the amplitude of density pertubations is closely related to
the power spectrum, which takes the form
$P(k)=|\delta(k)|^2\,D(z)^2$, where $\delta(k)$ gives the relative
strength of fluctuations on different scales, and $D(z)$, the growth
rate, describes the rate at which the amplitudes of fluctuations grow
with time.  We may approximate the growth of a cluster
as being self-similar with a characteristic mass $M_0$ that grows as
some power of the growth rate which depends on $|\delta(k)|^2$:

\begin{equation}					\label{mass}
M_0\propto D(z)^{\gamma}
\end{equation}

\ni where

\begin{equation}					\label{growthrate}
D(z) = \cases{ (1+z)^{-1}	&        if $\Omega=1$ \cr\cr
	1+{3\over x}+{{3(1+x)^{1/2}}\over{x^{3/2}}}\ln{[(1+x)^{1/2}-x^{1/2}]}
			&        if $\Omega<1$}
\end{equation}

\ni and $x(z) = \left|{1\over\Omega_0} - 1\right|\,{1\over{1+z}}$.

For $x<<1$, at high redshift, the growth-rate approaches that of the
Einstein-deSitter case.  However, at small redshift, or $x>>1$, the
growth-rate is approximately 1, after which point the cluster mass stops
growing with increasing time.  Thus if $\Omega<<1$, the mass of a
cluster is roughly constant after $z\approx 1/\Omega_0 - 2$, (or $z=3$
if $\Omega=0.2$, and $z=0.5$ if $\Omega=0.4$), while if $\Omega=1$ the
cluster mass continues to increase rapidly until the present day.

Over a wide range of scales, $|\delta(k)|^2$ can be approximated as a
power-law, $k^n$, for which $\gamma={6\over{n+3}}$, ranging from $2$
to $\infty$ for $n=0,-3$.  For the mass scales in which we are
interested, the cluster-cluster correlation function favors $n=-1.2$,
or $\gamma=3.3$ (Baugh \& Efstathiou 1993, Fisher et al. 1993, Vogeley
et al. 1992).  However, given that the technique described in the
paper observes only the highly non-linear cluster cores, we may expect
a different rate of evolution than predicted by linear theory alone.
Observations of large gravitationally lensed arcs and massive x-ray
halos in clusters at high redshift (Gioia \& Luppino 1994, Luppino et
al. 1994, Dickinson 1993) immediately suggest that there were already
massive cluster cores at $z\gta0.5$, and thus favor the smaller
values of $\gamma$.  Because of this uncertainty, in \S\ref{results}
we will not restrict ourselves to one particular value for $\gamma$,
and will instead consider models for a range of $\gamma$.

We now may use equation~\ref{mass} to calculate the characteristic
cluster luminosity $L_0$:

\begin{equation}
L_0(z) = L_0(z=0)\,D_0(z)^{\gamma}\,\Upsilon(z)
\end{equation}

\ni where $D_0(z)$ is the appropriate form for $D(z)$, normalized to be
1 at $z=0$, and where $\Upsilon(z)$ absorbs the evolution of the
cluster mass-to-light ratio due to galaxy luminosity evolution
(equation~\ref{upsilon}).  $L_0(z)$ absorbs the entire evolution of
the cluster luminosity function by simply shifting $n_L(L,z=0)$ towards
smaller luminosity with increasing redshift, without changing shape.

%============================

It is better to recast $N(\Sigma,L,z)$ (equation~\ref{Nsigmalz}) in
terms of directly observable quantities.  To do so, we must transform
the cluster luminosity $L$ into the angular size $\theta$ of the
cluster.  The size of the cluster is set by the radius at which the
surface brightness of the cluster falls below some limiting isophote,
and is therefore sensitive to the surface brightness profile of the
cluster.  We assume that the central regions of the cluster
are fit by a Hubble profile at $z>0$, which allows us to calculate the
apparent angular size of the cluster\footnote{This is a strong
assumption.  If clusters grow through continuous merging of
sub-clusters of comparable mass, this picture of simple self-similar
growth is an approximation.  N-body simulations can possibly test
these assumptions, and may easily be incorporated into further
refinements of the analytic work presented in this paper; however it
is unclear if they have either sufficient resolution or physics to
correctly model the evolution of the core.  Simulations could also be
used to directly measure the contribution of distant clusters to
surface brightness fluctuations.}.  Note that by assuming that the cluster
surface density is invariant while the cluster mass is changing, we
are implicitly assuming that the core radius of the cluster grows with
time.  Taking $\Sigma_l$ to be the surface brightness at the limiting
outer isophote, and $d_a(z)$ to be the angular diameter distance at
redshift $z$, the apparent angular size of the cluster is

\begin{equation}					\label{theta}
\theta =
{{r_c}\over{d_a(z)}}\,\left[\left({\Sigma\over{\Sigma_l}}\right)^{1/2}-1\right]
\end{equation}

\ni or

\begin{equation}
\theta = (41.3^{\prime\prime}) \, \left({{r_c}\over{0.1\Mpc}}\right) \,
\left({{10^3\Mpc}\over{d_l(z)}}\right) \,
\left({\left({\Sigma\over{\Sigma_l}}\right)^{1/2}-1}\over{2}\right).
\end{equation}

\ni which immediately indicates that fluctuations in the extragalactic
background light created by high-redshift, unresolved clusters
will easily fit within the field of view of most detectors.
The physical size of the cluster, $R_l$, at the limiting isophote is

\begin{equation}					\label{Rlim}
R_l = r_c \, \left[\left({\Sigma\over{\Sigma_l}}\right)^{1/2}-1\right].
\end{equation}

\ni Therefore, a cluster whose apparent central surface brightness
$\mu$ is 2.5 magnitudes brighter than the limiting isophote $\mu_l$
will be detected out to roughly two core radii.  We should also remark
at this point that, as discussed in \S\ref{coma}, the limiting isophote is in
fact a function of $\theta$, and would more appropriately be written
$\mu_l(\theta)$.  Including this effect would certainly complicate the
machinery being developed in this paper, and in practice may be
sidestepped by either analyzing a sample of clusters using a fixed
$\mu_l$ set by the limiting isophote of the smallest scale on which
clusters were searched for, of by redoing the analysis for several
different ranges of $\theta$, using an approximately correct $\mu_l$
for each, and then binning the results.

We may now relate $\theta$ to the luminosity using equation~\ref{theta}
and

\begin{equation}					\label{l_abellequ}
L(\Sigma,\theta,z) = 8\pi^2 r_c^2(\Sigma,\theta,z) (1+z)^4 \Sigma
\left[\ln{(1+x)} - {x\over{1+x}}\right],
\end{equation}

\ni the luminosity contained within a radius $xr_c$ for a Hubble
profile, where

\begin{equation}
x \equiv \sqrt{{{4\pi(1+z)^4\Sigma}\over{\Sigma_{lum}^\prime}}} - 1.
\end{equation}

\ni Equation~\ref{Rlim} corresponds to the luminosity
used in the Moore, Frenk, \& White luminosity function
(equation~\ref{n(L,z=0)}).  Figure~\ref{clustproppic} shows $L$ as a
function of apparent angular size $\theta$ for several values of the
apparent central surface brightness $\Sigma$.  Note the peculiar fact
that at a fixed apparent size and redshift, a cluster with a brighter
central surface brightness in fact has a smaller total luminosity.
This reflects that the cluster has a smaller effective radius
(see Equation~\ref{Rlim}), and thus a more steeply declining surface
brightness profile than a cluster of the same apparent size but a
fainter central surface brightness.

With equation~\ref{l_abellequ}, we may change variables in
$N(\Sigma,L,z)\,d\Sigma\,dL\,dz$ to get
$N(\Sigma,\theta,z)\,d\Sigma\,d\theta\,dz$, the number of clusters per
sterradian with apparent central surface brightness $\Sigma$, apparent
angular size $\theta$, and redshift $z$.  This yields

\begin{equation}					\label{N_z}
N_z(\Sigma,\theta,z) \, d\Sigma \, d\theta \, dz
	=  {n_L(L(\Sigma,\theta,z),z)\,\left|{{dL}\over{d\theta}}\right|\,
     d\theta\,dV(z)} { \,\times\, p(\Sigma,z)\,d\Sigma}
\end{equation}

\ni where

\begin{equation}
\left|{{dL}\over{d\theta}}\right|	= {{2L(\Sigma,\theta,z)}\over{\theta}}
\left[1-{1\over2}\left({x\over{1+x}}\right)^2\,\left({1\over{\ln{(1+x)} -
{x\over{1+x}}}}\right)\right]
\end{equation}

\ni Equation \ref{N_z}  may be integrated over all redshift to yield
$N(\Sigma,\theta)$:

\begin{equation}				\label{n_sig_theta}
N(\Sigma,\theta)\,d\Sigma\,d\theta = {\int^\infty_0 N_z(\Sigma,\theta,z)\,
				d\Sigma\,d\theta\,dz}.
\end{equation}

\ni This may also be recast in terms of $\mu$ using
equation~\ref{mufromsigequ}.

%----------------------------------------------

\section{Detection Rates and the Distribution of Surface Brightnesses}
\label{results}

Integrating Equation \ref{n_sig_theta} over $\theta$ and switching
variables from $\Sigma$ to $\mu$, we may calculate the distribution of
apparent central surface brightnesses, $N(\mu)$.  The distributions
are plotted in Figure~\ref{n_mu_pic} for a variety of cosmological models.
We have calculated $N(\mu)$ for $\Omega_0=0.2,1$, $H_0
=50,100\,\hnot$, $\gamma=0,1,2,3$, and for three different
assumptions about the spectral evolution of galaxies in the cores of
clusters.  In one case the cores of clusters are assumed to be
entirely populated by elliptical-like galaxies formed at $z=4$.  In
the second case the cores are filled with spiral-like galaxies formed
at $z=2$.  In the final case, which can more readily used to compare
the effects of varying the different cosmological parameters, we
assume that there is no evolution in the spectra of the cluster
galaxies, and we ignore the change in effective bandpass with redshift.
Hopefully the most
appropriate model for galaxy evolution lies somewhere in the space
spanned by the models used here.  (The models are described in more detail
in \S\ref{theory}.)  We have assumed a limiting isophotal magnitude of
29 mag/arcsec$^2$.

There are a few general points to note about the resulting
distributions:

(1) The detection rates can be almost arbitrarily high or low,
depending on the adopted form for the cluster mass and luminosity
evolution, and on the values of cosmological parameters.  Surface
densities of tens of clusters per square degree seem common, but can
be as high as several hundred clusters per square degree, and as low
as one cluster per ten square degrees.

(2) The distribution of surface brightnesses is highly peaked in
$\mu$.  The peak of the distribution tends to fall between 25 and 28
mag/arcsec$^2$ (for $\mu_{lim}=29$).  If the limiting isophote is
moved to a brighter level, the same angular size selects groups and
clusters that would appear at larger angular sizes for the fainter limiting
isophote.  See the next point for a discussion of how this affects
$N(\mu)$.

(3) The detection rates are smaller for fluctuations with large
angular size.  This reflects two fairly simple facts.  First,
intrinsically large systems tend to have larger angular extents;
therefore, at large angular sizes one is picking up richer systems,
which are rarer and have a correspondingly lower number density.
Second, because clusters that are nearby will have larger angular
sizes than more distant ones, by restricting a survey to large angular
sizes, one is probing a more local, smaller volume of the universe;
this again leads to a smaller surface density on the sky.

(4) Because the
systems selected at large angular size are closer on average, they
have also suffered less cosmological $(1+z)^4$ dimming; therefore, the
distribution of $N(\mu)$ peaks at brighter surface brightnesses for
large angular sizes.

(5) The surface density of clusters increases for smaller $\Omega$
and larger $H_0$.  For small $\Omega$, there is both a larger accessible
volume and a reduced rate of cluster evolution; this leads to larger
surface densities of richer systems.  For larger values of $H_0$, the
luminosity distance is smaller, which makes clusters visible out to
larger distances, increasing the accessible volume.

(6) The effects of galaxy luminosity evolution and cluster mass
evolution, (parameterized by $\Upsilon(z)$ and $\gamma$, respectively)
are very strong and are the largest uncertainties in the determination
of the surface density of fluctuations.  As will be shown in
\S\ref{applymethod}, clusters must be at redshifts greater than
$\approx0.5$ before their properties enter the range of surface
brightness and angular size to which surveys are likely to be
sensitive.  If there is rapid evolution of the mass of cluster cores
(i.e. large $\gamma$), then at these high redshifts there
will be so little mass assembled in what will eventually be the
cluster core that the cluster will be too faint to be detected.  However, if
the galaxies are very young and actively star-forming at high
redshifts, then the cluster luminosity and surface brightness can be
greatly enhanced by the increased star-formation.  Therefore strong galaxy
evolution could completely cancel the effects of the cluster
evolution.  Disentangling these two competing effects will be quite
difficult until information is brought to bear from other observations
of distant clusters and galaxies.  The masking of one effect by the
other can limit the usefulness of this method in constraining
cosmological scenarios.  If one is brave enough to trust one's models,
there is perhaps some differentiation that could be extracted from the
location of the peak surface brightness; we consider the required
level of trust to be premature at this time.

\section{The Redshift and Richness Dependence of $N(\Sigma,\theta,z)$}
\label{results2}

We may use the distribution of $N(\mu,\ln{\theta},z)$ and
$L(\Sigma,\theta,z)$ to understand the redshifts and richnesses of the
clusters that contribute to $N(\mu)$.  Both of these functions are
shown in Figure~\ref{n_mu_theta_z_pic}.  First note that for a given
surface brightness $N(\mu,\ln{\theta},z)$ tends to be peaked at a
particular redshift.  In the models that include the effects of galaxy
evolution, the distribution is much broader in redshift; at high
redshift, the cluster galaxies are much younger and are postulated to
have much higher star formation rates, and correspondingly larger e+k
corrections, which help to compensate for the decrease in flux at
large distance.  These models suggest that the clusters found could
exist at redshifts between 0.5 and 3.

Second, notice the strong effect that increasing $\gamma$ has on pushing
$N(\mu,\ln{\theta},z)$ owards smaller redshifts.
In models which assume that the
typical cluster mass evolves rapidly with redshift ($\gamma>0$), only
at $z<1$ are groups and clusters large enough to produce significant
fluctuations in the EBL.  The redshift distributions of clusters found
in the EBL could potentially be used to constrain the value of $\gamma$.

The top plots in Figure~\ref{n_mu_theta_z_pic} show how the luminosity
of clusters detected at a fixed angular size changes with redshift and
central surface brightness.  The luminosity is given as the luminosity
that the cluster would have {\it today}, after mass and spectral
evolution have taken place.  By comparing the top and bottom figures,
one can read off the richness of the system that is being detected at
a given redshift and surface brightness.  Not surprisingly, the
largest contribution to $N(\mu,\ln{\theta},z)$ usually comes from
poorer systems, which are more numerous.  We ignore contributions of
systems which have present day luminosities below $10 L^*$, the
luminosity of the smallest of Morgan's poor clusters (Bahcall 1980,
Albert et al. 1977, Morgan et al. 1975); below this luminosity, the
radial profiles of poor groups are not well understood, and are
probably not well defined for an individual group.  With the exception
of the dense, but rare, groups typified by Hickson's compact groups
(Hickson 1982), it is not unreasonable to assume that poor groups will
be less tightly bound and thus will have low central surface
brightnesses, and will be nearly undetectable when placed at a
distance such that their component galaxies cannot be seen
individually.

Figure~\ref{n_mu_theta_z_pic} also demonstrates how the
richness of a group or cluster that we detect in the EBL depends
on its redshift.  In the absence of galaxy evolution, a system that we
detect at high redshift will be richer than a system at lower redshift
which has the same observed central surface brightness and angular
size.  This is due to the larger $(1+z)^4$ cosmological dimming that
sources at high redshift experience, as well as the (typically) larger
physical size that corresponds to a given apparent angular size
(e.g. Equation~\ref{theta}).  If galaxy evolution is included, however,
a poor cluster of young galaxies at high redshift may have the same
apparent size and surface brightness as a richer, older cluster at
lower redshift.

%----------------------------------------------
\section{Some Thoughts on Applying this Method}		\label{applymethod}

We now turn our discussion from
rather idealized theoretical concerns to the more grisly, messy
issues related to carrying out an actual survey that is optomized
to detect the richest systems.

As discussed above, the EBL must show signatures of both distant rich
clusters and nearer, poorer groups.  The degeneracy that exists
between distant rich systems and nearby poor systems is a difficult
one to break, especially when all objects of interest are at
high enough redshift that the angular diameter distance changes very
little.  However, there are some steps one may take to maximize
the efficiency of finding distant rich clusters.

First, richer clusters dominate at lower surface brightness, for a
fixed angular size.  This may at first seem counterintuitive; however,
a system with a radial surface brightness distribution with a low
central surface brightness and a large angular size has a
more slowly falling profile, and thus a greater integrated luminosity
beyond the central core, than a more centrally peaked system.
Therefore, by restricting one's attention to fluctuations with lower
surface brightness, one is more likely to be selecting for richer
systems.

Second, we may maximize the contribution from rich clusters by
selecting objects of a particular angular size that have few
individual galaxies within the area of the EBL fluctuation.  The
expression for $L(\Sigma,\theta,z)$ given in equation~\ref{l_abellequ}
shows that the rate at which $L$ rises with redshift strongly
increases with $\theta$; this simply reflects the rather simplistic
rule that nearby objects tend to look larger than distant ones.
Fluctuations seen at large angular size will therefore tend to be
closer.  (As an example of this, for an angular radius of $30\arcsec$
and a central surface brightness of $26\surfb$, a cluster with
$L_{1.5}=100L^*$ would have these properties at $z=1.5$ in a
no-evolution model, but a similar cluster must have the much
larger redshift of $z=2.5$ if the detected angular radius were reduced by a
factor of 2.  A group with the minimum luminosity $L_{1.5}=10L^*$,
would appear at $z=1,0.5$ for $\theta=15\arcsec,30\arcsec$,
respectively.)  By properly choosing the angular size, poorer clusters
will be brought close enough that the individual galaxies within the
cluster could be readily seen, while richer clusters, which tend to be
futher away (for the same angular size and smoothed surface
brightness), will have fewer of their component galaxies above the
magnitude limit of the survey.

How close must a cluster be before one could see its component
galaxies?  We can estimate the redshift beyond which the cluster
receeds into the background light in two ways.  First, we can estimate
the apparent $m^*$ of the cluster as a function of redshift.  A
cluster that is far enough away that its $m^*$ drops below the
magnitude limit of the survey will have too few members visible to be
detectable as a cluster in the traditional way.  The unsmoothed survey data
would yield little information on the richness of the cluster.
Assuming that $M^*=-21$ and a
magnitude limit of $m_{lim}\gta 23$, $L^*$ galaxies in the cluster
should be visible at redshifts less than 0.5-1, depending on the
details of the evolution.

For a more empirical limit on the redshift at which cluster galaxies
become visible, we can look at the apparent redshift of the brightest
cluster galaxy (BCG) as a function of redshift and richness.  (For
simplicity, we will ignore the misnomer and take ``BCG'' to mean the
brightest galaxy in a cluster {\it or} group).  To determine the
apparent magnitude of a BCG with redshift, we construct a $V$-band
Hubble diagram from the high redshift ($0.5\lta z\lta0.9$) clusters in
Arag\'on-Salamanca et al. (1993), the intermediate redshift ($0.2\lta
z \lta 0.4$) clusters in Pickles \& van der Kruit (1993), and Abell
clusters from Schneider, Gunn, \& Hoessel (1983) (See
Figure~\ref{BCGpic}).  We fit the inverse Hubble diagram for rich
clusters with two straight lines

\begin{equation}
z_0(V) = \cases{0.0471\times V - 0.6587 & if $V\lta18$ (roughly $z\lta0.2$) \cr
		0.1423\times V - 2.3260 & if $18\lta V$ (roughly $0.2\lta z\lta1$)
		}.
\end{equation}

As very little is known about clusters or BCG's above $z=1$, it is
difficult to extend this relation to higher redshifts.
However, radio galaxies, which are often identified as BCG's,
have apparent $V$ magnitudes of $\approx22.5$
at $z>1$ (Djorgovski, Spinrad, \& Dickinson 1987).
If true, then the presence or absence of foreground
galaxies is a poor distance discriminant beyond $z\approx1$.  Thus, an
angular size should be chosen such that the typical redshifts of the
poorest systems one is interested in lie below $z=1$.

To extend the apparent magnitude of BCG's to poorer systems, we have
estimated the magnitude of the brightest group galaxy as a function of
group size using the CfA Group Catalog (Geller \& Huchra 1983).  The
constancy of $M_{BCG}$ with richness seems to extend to clusters with
a luminosity of $\sim 0.2$ of the luminosity of the Coma cluster
(corresponding to roughly 30 group members in the CfA $m<14.5$
Catalog).  We define this luminosity to be $L_c$.  For poorer groups,
the typical BCG becomes fainter, as would be expected if merging in
the high density cores of clusters is resposible for enriching large
central CD galaxies.  A reasonable empirical fit to the change in
$M_{BCG}$ with group luminosity is, relative to the typical absolute
magnitude of a BCG in a rich cluster,

\begin{equation}
\Delta M_{BCG}(L) \approx \cases {
   2.4\left(1 - {{L}\over{L_c^*}}\right) &if $L<L_c$\cr
   0						&if $L\ge L_c$\cr
   }.
\end{equation}

\ni We fix the value of $L_c$ from Abell's luminosity of
the Coma cluster as used by Bahcall (1979) and find $L_c\approx20L^*$.
There is considerable scatter in this relation, but the general trend
of groups having fainter BCG's than rich clusters seems evident.  This
shifts the lines in Figure~\ref{BCGpic} towards smaller redshifts.

%----------------------------------------------
\subsection{Other Sources of Fluctuations}

Distant clusters will of course not be the only source of fluctuations
in the EBL.  Any source which causes low level surface brightness
fluctuations in the background light will contribute to the observed
$N(\mu)$.  We are fortunate, however, that many of the possible
sources are either negligible or easily separated.  Preliminary
results from a survey for nearby low-surface brightness galaxies
(Dalcanton 1995), as well as follow-up observations at Kitt Peak, show
that large LSB's have a very low surface density on the sky,
particularily at the faint surface brightnesses that are relevant for
detecting clusters.  Therefore, they will not be a substantial source
of confusion.  Likewise, low-surface brightness extensions to
disturbed galaxies, such as might be due to tidal interactions or
merging, will be fairly easily distinguished by the presence of the
disturbed galaxy adjacent to the detected fluctuation.  Other
instrumental sources of low-surface brightness fluctuations, such as
artifacts due to scattered light and flat fielding, are not fixed to
the sky, and thus can be separated by moving the telescope; for
example, repeating a drift scan with slight shifts in declination
would be one way of constraining instrumental contributions.
Scattering and flat-fielding errors are usually either linear features
or large scale variations, both of which will have signatures that are
distinct from the ``bumps'' due to distant clusters.  The one
remaining source of fluctuations that could potentially be confused
with clusters is chance superpositions of physically unrelated faint
field galaxies.  In the Appendix we show that the
contribution from a uniform background of field galaxies is small,
especially on the large angular scales over which we are interested.
Typically the Poisson fluctuations of the EBL itself dominate the
fluctuations from the field galaxies.  This is supported by the analysis of
MS1054 presented in \S\ref{coma}.

%----------------------------------------------
\section{Conclusions}

The technique of looking for high-redshift clusters manifested as
low-surface brightness fluctuations in the background light
presents a powerful new tool for finding one of the most elusive of
astronomical objects.  The method we have proposed could isolate
high-redshift clusters at a significant rate, up to hundreds per
square degree, depending on the particular form of cosmology and
galaxy evolution that Nature has chosen.  Clusters give us a way of
isolating a population of galaxies at a particular redshift, and
therefore finding the highest-redshift clusters opens a window onto
some of the earliest stages of galaxy evolution.  The surface density
of clusters that this method could find would also be a constraint on
various scenarios for structure formation.  More accurate simulations
of cluster profiles as a function of redshift in different cosmologies
and more work on untangling luminosity evolution from mass evolution
are required before this method could provide a strong constraint, however.

Looking for clusters in the EBL is a method that we have shown works
in principle, and in practice.  The technical limitations involved are
few and readily surmountable.  The extremely accurate flat-fielding
required is easily reached with telescopes operating in drift-scan
mode.  And while searching for low-surface brightness fluctuations
requires extremely accurate flat-fielding, it has dramatically less
stringent requirements on exposure time than traditional direct
imaging.  This means that this method could easily survey large areas
of the sky in just a few nights.  Deeper follow-up imaging would be
required to confirm the existence of clusters.  As more is
learned about the composition of clusters at higher redshift, color
information could be incorporated into the cluster selection criteria
to increase the efficiency of cluster selection.  A region of the sky
could be observed through multiple band passes, and the colors of the
bumps in the EBL could be used to help isolate high-redshift clusters,
and perhaps shed light on the global behavior of star formation in
galaxies at redshifts greater than 1.  Similarily, information from
other wavelengths could be incorporated.  Regions of large
fluctuations in the optical EBL could be compared to the smoothed
x-ray EBL or to catalogs of radio galaxies to increase the likelihood of
identifying rich systems.

%----------------------------------------------
\bigskip
\bigskip
\centerline{Acknowledgements}
\bigskip

The idea for this paper grew out of a series of ongoing discussions
with my thesis advisor, David Spergel.  To
him goes much of the credit for encouraging me to pursue a nutty idea
and turn it into something vaguely sensible.  (If it does in the end
turn out to be nutty, however, I accept full credit.)  It is a pleasure
to thank Gerry Lupino and Isabella Gioia for giving me access to their
lovely image of MS1054, which provided the first definitive test of
the methods proposed in this paper.  I also thank Gerry for sharing
his image display secrets.
Jim Gunn and Dan Rosenthal are
also warmly acknowledged for their encouragement and advice,
as is Teresa Shaw for assistance with typing.

\bigskip
\bigskip
\centerline{\bf Appendix A}
\centerline{\bf Contributions of Faint Field Galaxies}
\bigskip

Assuming a power law distribution in magnitude for the faint galaxy
number counts, we may use Monte Carlo simulations to estimate the
surface brightness at which faint field galaxies will begin to
contribute significantly to the fluctuations.  A high surface
brightness fluctuation in the background of field galaxies will be
sufficiently rare to make effectively no contribution to the
number of fluctuations.  At lower surface brightnesses, however,
fluctuations due to field galaxies will be more common, and will thus
give an appreciable signal.  The surface brightness threshold at which
this happens will be a strong function of the scale length over which
the background sky is smoothed.  If the sky is smoothed over large
apertures, then many field galaxies will be averaged together, and the
level of fluctuations will be smaller than if a small aperture were
used.

The aim of this section is to estimate the surface brightness limit
(as a function of smoothing length) at which background field galaxies
are expected to make a significant contribution to the number of
fluctuations.  The method for doing so is rather straighforward.
First, assume that the number of galaxies per magnitude per area
on the sky can be described by a single power law

\begin{equation}				\label{numcounts}
\log{\left( {{dn(m)}\over{dm\,d\Omega}} \right)} = \alpha(m - m_0),
\end{equation}

\ni as is seen in deep surveys of field galaxies (see Tyson 1994 for a
review).  Second, choose $N$ galaxies with magnitudes between $m_{min}$ and
$m_{max}$,

\begin{eqnarray}
N &= & \Omega \int^{m_{max}}_{m_{min}} {{dn(m)}\over{dm\,d\Omega}} \, dm \cr
  &=  & {\Omega\over{\alpha\,\ln{10}}}
	\left[{{dn(m_{max})}\over{d\Omega}} -
                                           {{dn(m_{min})}\over{d\Omega}}\right]
\end{eqnarray}

\ni where $m_{min}$ is the magnitude of the faintest galaxy
that {\it cannot} be detected above the pixel-to-pixel noise of the
sky, $m_{max}$ is the magnitude at which the faint galaxy number
counts fall off from a power law, and $\Omega$ is the area of the
aperture over which the sky is smoothed (i.e. the area of the
smoothing kernel).  Third, multiply the flux from each galaxy by a
value drawn from the two-dimensional smoothing function used within
the aperture, and then add the fluxes together to find the surface
brightness.  (The units of the smoothing function are
``probability/area'', so multiplying the fluxes by the smoothing
function automatically yields a surface brightness).  Finally, repeat
these steps many times to generate the probability distribution of
surface brightness fluctuations due to field galaxies.

We have performed this exercise using simple gaussian windows of
different sizes, truncated at $5\sigma$, and for a range of
assumptions about the slope of the faint galaxy number counts and
possible survey magnitude limits.  We have taken 1 ``galaxy clump''
per square degree to be the fiducial limit at which the contribution
from field galaxies becomes significant.  We find the surface
brightness which corresponds to this detection rate by first assuming
that we can treat a square degree of the sky as being $M=(1
{\rm deg}^2)/\pi\sigma^2$ independent regions and then finding the
surface brightness at which there is only a $1/M$ chance of reaching a
brighter surface brightness due to chance fluctuations in the surface
density of field galaxies.  This is obviously not strictly correct
given that neighboring regions of the sky are not independent.  It is
also not correct for very small smoothing lengths, where the
background galaxies cannot be treated as point sources.  However, it
is does serve as an estimate of the relevant surface brightness at which
we might expect field galaxies to become important.

The results are shown in Figure~\ref{eblpic}.  The contribution of field
galaxies to the fluctuations in the smoothed sky background is a
strong function of smoothing length, as expected, and becomes
effectively negligible for smoothing lengths greater than
$10^{\prime\prime}$.  At smaller smoothing lengths, the field galaxies
make a significant contribution at brighter surface brightnesses
($\mu\approx26.5$); however, as shown in Figure~\ref{eblpic}(b), the angular
size at a limiting isophote $\mu=29$ is never greater than
$10^{\prime\prime}$.  Therefore, if a survey is limited to smoothing
lengths greater than $10^{\prime\prime}$, or to finding objects with
angular size greater than $10^{\prime\prime}$, then we would expect
the contribution from faint field galaxies to be small.  These results
will of course change if different functions are chosen for the
smoothing window, and should be repeated depending on the details of a
particular survey.

%----------------------------------------------

\clearpage
%--------------------------------------------------------------

\newcounter{figcnt}

\begin{list}
{Figure \arabic{figcnt}.}  {\usecounter{figcnt}}

\item  The mean surface brightness within $\theta$ for a CD galaxy
(light solid line) and a Coma-like cluster (dark solid line) at
$z\approx 1.5$, neglecting luminosity evolution or k-corrections.
Also plotted is an example of the limiting surface brightness within
an aperture of radius $\theta$.  Note that while there is no aperture
size within which the mean surface brightness of the CD galaxy rises
above the limiting surface brightness, for large enough apertures the
mean surface brightness of the cluster is high enough to be
detectable.  The CD galaxy is taken to have a deVaucouleur's profile
with $r_e=35\kpc$ and a total luminosity of $4\,L^*$.  The cluster
has a Hubble profile with $r_{core}=0.1\Mpc$ and a luminosity within
an Abell radius (1.5 Mpc) of $L_{Abell}=100\,L^*$.  ($H_0=75$ km/s/Mpc,
$\Omega=0.2$)
\label{detectabilitypic}

\item  Detecting an artificial cluster with $L_{Abell}=90L^*$ and
$r_c=0.1Mpc$ at $z\approx1.5$ ($m^*=23.5$ and $\theta_c=15\arcsec$).  (a) An
artificial cluster of galaxies distributed in a Hubble profile with
$\theta_c=15\arcsec$ and with magnitudes drawn from a
Schecter-function with $m^*=23.5$ and $\alpha=-1.25$, down to $m=29$.
The entire image is $11.5\arcmin$ across.  (b) The same cluster from
(a), but embedded in a background of field galaxies with magnitudes
drawn from a power-law distribution ($\log{{dN}\over{dm\,d\Omega}} =
0.38(m-12.13)$) between $m=22.5$ and $m=29$.  (c) The cluster and
field galaxies from (b) embedded in Poisson noise with a surface
brightness of $21.3\surfb$.  Note that the galaxies all fall below the
detection limit of the image.  (d)  The image from (c), smoothed with
a Gaussian filter with $\sigma=13.5\arcsec$.  While there is only the barest
hint of the cluster in (c), the smoothed image shows the presence of
the cluster quite clearly due to the dramatically lower pixel-to-pixel
variation of the noise.  The peak is $9\sigma$ above the sky level.
\label{fakeclusterpic}

\item  Recovering a known high-redshift cluster using surface brightness
fluctuations.
(a) A 3-hour $I$-band image of the rich x-ray selected cluster MS
1054.5-0321 ($L_x = 9.3\times10^{44}$ ergs/s) at $z=0.823$, taken with a
Tektronics 2048x2048 CCD ($0.22\arcsec$ pixels in $0.8\arcsec$ seeing) on
the UH 88-inch telescope, kindly provided by Gioia \& Luppino (see
Gioia \& Luppino 1994).  The image is a composite of 18 600s images,
each flattened with a ``supersky flat'' made from the median of 50
deregistered images.  The typical shift between exposures
was $10\arcsec-20\arcsec$.  There are some obvious low-level, large-scale
flat-fielding variations across the image that are visible at much
higher contrast.
(b) The same cluster, but with Poisson-noise added to reduce the effective
exposure of the image by a factor of 400, producing an effective exposure of
27 seconds.  Note that almost all of the cluster galaxies now fall below
the level of the sky noise.
(c) The image from (b) ``cleaned'' of foreground objects.  FOCAS was used
to identify all $4\sigma$ peaks.  Bright, large objects were masked out
and replaced with the median sky level of the entire image.  The remaining
objects were replaced with values drawn from the sky histogram of a
ring around the object.  This preserves large low-surface brightness
features that lie below compact high-surface brightness features.  The
procedure used to generate this image is fully automated and is being
used for a survey of low surface brightness galaxies.  It is described
more fully in Dalcanton (1995).
(d) The image from (c), smoothed with a Gaussian filter with $\sigma=7\arcsec$.
The cluster is easily detected as the brightest and largest peak
on the image, lying $6\sigma$ above the sky with a diameter of roughly
$30\arcsec$.  Extended light from the
foreground group to the right of the cluster is also apparent, though
at lower significance.  There are also indications of some of the
principle sub-clumps at low significance as well.  The largest negative
deviations are only $3\sigma$ below sky, and correspond to regions of
obviously flat-fielding errors on the original image.  The level of
these deviations would presumably be smaller in a drift-scanned image,
increasing the significance of the cluster detection.
\label{emssclusterpic}

\item  The luminosity and bandpass evolution corrections used in the paper for
``elliptical'' and
``spiral'' galaxies (Equation~\ref{upsilon}.  The corrections are derived from
the models
described in Bruzual \& Charlot (1992), using
$z_{formation}\approx2,4$ and $t_{formation}/\tau=2,14$ for the
``spiral'' and ``elliptical'' galaxies, respectively, as viewed
through a Gunn $g$ filter.  Adding $-2.5\log{\Upsilon(z)}$ to
$5\log{d_l(z)}$ gives the full e+k correction.
\label{epluskpic}

\item  The properties of a cluster at $z=1.5$ as a function of apparent
angular size and central surface brightness.  (a) The luminosity
within an Abell radius ($1.5\Mpc$) in units of $L^*$ as a function of
apparent angular size $\theta$.  The lines are for different apparent
central surface brightnesses: $\mu=26$ ({\it solid}), $\mu=27$ ({\it
dotted}), $\mu=28$ ({\it dashed}), assuming $M^*=-21$.  See
equation~\ref{l_abellequ}.  (b) The core radius of the cluster as a
function of apparent angular size.  The lines correspond to the same
central surface brightnesses as in (a).  See equation~\ref{theta}.
\label{clustproppic}

\item  $N(\mu)$ for various parameters, in units of
\#/degree$^2$/mag/arcsec$^2$.  The solid and dashed lines
are $\Omega=1$ and $\Omega=0.2$, respectively.  The heaviest line is
$N(\mu,\theta)$ integrated between $10\arcsec$ and $20\arcsec$, the
medium weight line is $N(\mu,\theta)$ integrated between $20\arcsec$
and $30\arcsec$, and the lightest line is $N(\mu,\theta)$ integrated
between $30\arcsec$ and $40\arcsec$.  Note that the number of clusters
goes up for lower $\Omega$, smaller angular sizes, and smaller values
of $\gamma$.  Changing $\gamma$ affects the number of cluster more
dramatically for $\Omega=1$ than for low values of $\Omega$.  The
limiting isophotal magnitude is taken to be $\mu_{lim}=29$, and the
limiting redshift is taken to be $z_{lim}=2$.  Only groups that will
have luminosities greater than $10\,L_*$ at $z=0$ are included.
$h_{100}=1$ unless otherwise noted.
\label{n_mu_pic}

\item  $N(\mu,ln\theta,z)$ (in units of
\#/degree$^2$/mag/arcsec$^2$/ln(radians))
and the corresponding luminosity that the detected cluster will
eventually have at $z=0$.  These pairs of plots can be used to understand
the redshifts and richnesses of the systems that contribute to
surface brightness fluctations in different cosmological scenarios.  For
all of these plots, $h_{100}=1$ and $\Omega=1$.
(a) Models with
no luminosity evolution.  The different line types correspond to
different values of $mu$:
$\mu=24$   (heavy solid);
$\mu=24.5$ (heavy dot-dashed);
$\mu=25$   (heavy long dashed);
$\mu=25.5$ (heavy short dashed);
$\mu=26$   (heavy dotted);
$\mu=26.5$ (light solid);
$\mu=27$   (light dot-dashed);
$\mu=27.5$ (light long dashed);
$\mu=28$   (light short dashed);
$\mu=28.5$ (light dotted).
The columns show models with no cluster mass evolution ($\gamma=0$) on the
left, progressing to more dramatic cluster evolution ($\gamma=3$) on the
right.  While for a fixed $\mu$, $\theta$, and $z$, one always detects
a system with the same luminosity, for larger $\gamma$'s the
system will accrete more mass at late times than if $\gamma$ were smaller;
this effect can be seen as a gradual steepening of $L(\mu,\theta,z=0)$
with increasing $\gamma$.  Because the current number density of
clusters is fixed, if $L(\mu,\theta,z=0)$ grows more steeply with redshift,
then there will be fewer high redshift systems contributing to
the number of surface brightness fluctuations; this can be seen in the
behavior of $N(\mu,ln\theta,z)$, shown in the lower of each pair of plots.
The top row of plots shows $N(\mu,35\arcsec,z)$, which can be compared
with the integrated function $N(\mu)$ shown as the lightest lines
in Figure~\ref{n_mu_pic}).
The bottom row of plots shows $N(\mu,15\arcsec,z)$, which can be compared
with the darkest lines in Figure~\ref{n_mu_pic}).
(b) Models with ``elliptical'' luminosity evolution.
(c) Models with ``spiral'' luminosity evolution.
\label{n_mu_theta_z_pic}

\item  $V$ magnitude of the brightest cluster galaxy as a function of
redshift for three different samples.
\label{BCGpic}

\item  (a) The central surface brightness at which the fluctuations due
to a Gaussian-smoothed uniform background of field galaxies
contributes one peak per square degree, as a function of the size of
the Gaussian window.  The symbols correspond to different assumptions
for the magnitude of the brightest undetected galaxy and for the slope
of the faint number counts.  (b) The same central surface brightness
as in (a), but plotted as a function of the angular size where the
surface brightness fluctuation falls below a limiting isophote of
$\mu_{lim}=29$.  While at small angular sizes the contribution from
randomly distrubuted field galaxies to surface brightness flucuations
in the background sky can be significant, for angular sizes greater
$10\arcsec$ in radius, the contribution is negligible.
\label{eblpic}

\end{list}

\vfill
%----------------------------------------------
\end{document}